# Harnessing the Deep Web: Present and Future

Jayant Madhavan
Google Inc.
jayant@google.com

Loredana Afanasiev
Universiteit van Amsterdam
lafanasi@science.van.nl

Lyublena Antova
Cornell University
lantova@cs.cornell.edu

Alon Halevy
Google Inc.
halevy@google.com

## 1. INTRODUCTION

The *Deep Web* refers to content hidden behind HTML forms. In order to get to such content, a user has to perform a form submission with valid input values. The name Deep Web arises from the fact that such content was thought to be beyond the reach of search engines. The Deep Web is also believed to be the biggest source of structured data on the Web and hence accessing its contents has been a long standing challenge in the data management community [1, 8, 9, 13, 14, 18, 19].

Over the past few years, we have built a system that exposed content from the Deep Web to web-search users of Google.com. The results of our surfacing are now shown in over 1000 web-search queries per-second, and the content surfaced is in over 45 languages and in hundreds of domains. The algorithms underlying our system are described in [12]. In this paper we report some of our key observations in building our system and outline the main challenges we see in the further exploration and use of deep-web content.

To understand the different efforts on providing access to deep-web content, we first present the rapidly changing landscape of different kinds of structured data that exist on the web and the relationships between them (Section 2). In fact, there appears to be some confusion about the term Deep Web – it has often been incorrectly used synonymously with structured data on the Web. The Deep Web is one (significant) source of data, much of which is structured, but not the only one. We describe the different types of structured data in the context of the varying search tasks that we can strive to support over them.

Second, we discuss our choice of underlying approach in exposing deep-web content in a search engine. Most prior works on the Deep Web have espoused one of two main approaches. The first, known as virtual integration, follows the data integration paradigm. Here, we consider each deep-web site as a source in a data integration system. Users pose queries over a mediated schema that is exposed to them as a web form, and queries are routed to the relevant sites. The second, known as surfacing, attempts to pre-compute queries to forms and inserts the resulting pages into a web-search index. These pages are then treated like any other page in the index and appear in answers to web-search queries. We have pursued both approaches in our work. In Section 3 we explain our experience with both, and where each approach provides value.

We argue that the value of the virtual integration approach is in constructing vertical search engines in specific domains. It is especially useful when it is not enough to just focus on retrieving data from the underlying sources, but when users expect a deeper experience with the content (e.g., making purchases) after they found what they are looking for. However, the virtual approach does not scale well when building a single system to seamlessly include more than a small number of domains, while retaining a simple keyword querying interface. The challenges to be addressed include routing, i.e., identifying the (small) subset of sites that are likely to be relevant to a keyword query, and reformulation, i.e., translating the keyword query appropriately for each site. We note that the keyword reformulation problem is different from the query reformulation problem studied in the context of data integration systems.

In contrast, we argue that the surfacing approach is a better fit for web search where we are answering keywords queries that span all possible domains and the expected results are ranked lists of web pages. We find that the web pages we surface add significant value to search engine traffic and their impact is especially significant in smaller and rarer domains and queries. The challenges in surfacing arise in determining a set of queries for each form. Our results in [12] have shown that understanding the exact semantics of forms does not play a significant role in surfacing. However, we argue (in Section 4) that there is value in inferring some semantics, such as popular input data types that are prevalent across domains (e.g., zipcodes), and the correlations between inputs within forms (e.g., maxprice and minprice). Addressing these challenges can significantly improve the quality of surfacing.

Third, we discuss the challenges in manipulating the content that is exposed by surfacing (Section 5). We observe that since surfaced web pages are treated simply as text documents, the semantics of the underlying content is being ignored. Of course, losing the semantics of the surfaced content is a lost opportunity for query answering. Retaining and utilizing the semantics of exposed content are interesting problems for future research. We also note that it is interesting to understand the extent to which any surfacing approach exposes the contents of a deep-web site. This chal-lenge of estimating *coverage* is also relevant in the context of database exploration [1, 13] and remains an open problem.

Lastly, we note that while accessing individual sources of structured data on the Web is of great interest, there is also a significant opportunity to use the *aggregate* of this vast quantity of structured data and meta data. In particular, by analyzing collections of structured data we can obtain large collections of schemata, values for column labels and relationships between values. These resources can then be used to assist information extraction efforts and in the reformulation of web-search queries. We describe (in Section 6) our initial work on building a *semantic server* that provides several services that are built by analyzing large amounts of structured data on the Web.

## 2. THE LANDSCAPE OF STRUCTURED DATA ON THE WEB

As background for our discussion, we begin by describing the kinds of structured data on the Web today, and place the Deep Web in that context. One of the important points we make is that the Deep Web is just *one* type of structured data on the Web. (There is often a misconception that Deep Web and structured data on the web are synonymous). However, before we describe the different types, it is important to first consider the *search task* we expect to use the structured data for. Roughly, these structured data tasks can be classified into one of the following classes (note that we make no claim that this is a comprehensive list):

**Unstructured queries:** These tasks are identical to the popular current mode of searching for information on the Web. Users pose keyword queries and get a listing of URLs to web pages as the result. The goal here is to ensure that web pages that contain structured data get ranked appropriately high in result listing. A variation on this task is the search for structured data collections (i.e., return pages that contain HTML tables or mashups). Such a search may be invoked when one is collecting data for a mashup or to conduct a more detailed study of a phenomenon.

**Single-page structured queries:** In these tasks we pose more precise queries to the search engine. We may pose them using an interface that allows users to submit more structured queries (e.g., searching for jobs, cars or houses), or expect the underlying engine to parse a keyword query and recognize its structure (e.g., identify in the query "france population" that France is a country and population is a property of countries). There are two types of answers we can expect. The first is the precise answer (e.g., the actual number for the population of France). The second is a page that contains the precise answer (e.g., the Wikipedia page on France), but the user needs to read through the page to find the answer.

**Multi-page structured queries:** These tasks are more ambitious than the previous set. Here we expect the search engine to find answers that are derived by *combining* (e.g., via a join or union) data from multiple structured (or unstructured) sources on the Web. For example, we would like to combine data from the Internet Movie Database and web sites with movie playing times to find local playing times of movies directed by Woody Allen [10]. Note that the user need not be aware that the answers were obtained by combining data from multiple pages.

Given these tasks, we consider the structured data we find on the Web and find that it comes mainly in two forms: one where the data itself is already structured in the form of tables, and the second where the *query interface* to the data is structured. The first type of data was analyzed by the WebTables Project [3] that collected all the HTML tables on the Web, extracted the ones that offer hiqh-quality data, and offered a search interface over them. That work showed that there are on the order of 150 million high-quality relational tables on the Web when restricting ourselves to English pages. *Structure* here means that the tables have rows of data, and in many cases there is a row with names for the attributes (though in some cases inferring that such a row exists is non-trivial).

The second kind of structured data is available through structured query interfaces. HTML-form interfaces enable users to pose specific (but templated) queries in particular domains and obtain results embedded in HTML pages. However, these results are not as structured as the HTML tables mentioned above. Significant additional effort is required in order to extract tabular data from these results, and writing such extractors is often highly dependent on the specific site being targeted. Hence, information extraction on a large-scale from this data remains a significant challenge. The Deep Web is a subset of this second kind of data. It is only a subset because in many cases there are HTML links that lead to the same result pages, and therefore crawlers can reach them without additional machinery.

Returning to the tasks we mentioned above, deep-web data is clearly useful for unstructured queries and single-page structured queries that return web pages as results. In [12] we describe how we crawled millions of forms by pre-computing a set of relevant queries for each form. The web pages we extract from that crawl are inserted into the web search index and served like any other web page, thereby fulfilling the first kind of task. To enable single-page structured queries, one typically needs to restrict the application to a particular domain and create a repository of forms with descriptions of what data can be obtained from each one and then route queries to relevant forms at query time (this approach is referred to as the virtual-integration approach). However, providing precise answers from the results of deep-web content is significantly harder because of the additional extraction effort that is required and the fact that extraction is hard to do in a site-independent way for a large numbers of sites. In Section 3 we explain the merits and limitations of both of these approaches to accessing the Deep Web.

The WebTables data, on the other hand, has been used for unstructured queries where the goal is to find structured collections of data, and can be used for finding precise answers to single-page structured queries. One of the challenges in doing so is that while the column labels of HTML tables may be present (or detectable using techniques described in [2]), the semantics of the table (e.g., the table name) is often embedded in the text surrounding it and therefore harder to detect. For example, we may have a table with two columns, Name and Year, but the table can refer to the winners of the Boston Marathon or winners of the Oscars. We are currently investigating using the WebTables data for answering multi-page structured queries.

We note that there are several other kinds of structured data on the Web. The first is a growing number of mashups and other interactive visualizations that can now be created with several convenient tools. Such mashups display very valuable data, but the challenge in supporting queries on them is that they are often hidden behind Javascript code. Large collections of data created by mass collaboration such as Freebase, DBPedia, and Google Base are also valuable resources, though their coverage is much narrower than the Deep Web. The second kind is semi-structured data such as blog posts and discussion forums. In these contexts, the articles have meta-data associated with them (such as author, date, thread they are responding to), and this meta-data can be extracted with some effort. We anticipate that the tasks for which this structure will be useful are more specific, such as sentiment analysis and trend analysis on the blogosphere. Hence, the queries posed over this data will be highly specific and often embedded in application code. Finally, some structure is added to content via annotation schemes for photos (e.g., Flickr), videos (e.g., YouTube) and bookmarks (e.g., del.icio.us). It is easy to see how such annotations can be used to enhance unstructured queries and for refining the results of queries with larger result sets.

## 3. VIRTUAL INTEGRATION VERSUS SURFACING

As we described in the previous section, there area two main approaches to providing access to deep-web content: virtual integration and surfacing. The virtual-integration approach has been used in companies and in research typically under the umbrella of vertical search (e.g., [4, 16, 17]). We have pursued both approaches. In the past, two of the authors used the virtual-integration approach to implement a search engine for classifieds (e.g., used cars, apartments, personals, etc.) [5]. More recently, we have used the surfacing approach to include deep-web content into web-search at Google.com. Based on our experiences, in this section we explain the benefits and limitations of both approaches in practice.

### 3.1 The virtual integration approach

The virtual-integration approach is basically a data integration solution to accessing deep-web content. The approach is based on constructing mediator systems, potentially one for each domain (e.g., used cars, real-estate, or books). To create an application with this approach, one analyzes forms and identifies the domain of their underlying content, and then creates semantic mappings from the inputs of the form to the elements in the mediated schema of that domain. The mediated schema for each domain can either be created manually or by analyzing forms in that domain (e.g., [7, 15]). Queries over the mediated schema can then be reformulated as queries over each of the underlying forms. Results retrieved from each of the forms can potentially be extracted, combined, and ranked, before being presented to the user.

Virtual integration is a very attractive option when designing vertical search engines. Vertical search engines focus on particular (large) domains or on particular types of entities in a domain (e.g., used car listings, apartment rentals, real-estate listings, airline reservations). They improve user experience by enabling richer and more-specific queries, e.g., searching for cars by color, price, and year. Such queries are a better fit for domain-specific advanced search forms rather than simple keyword queries in search engines. The form-selection and reformulation problems are more manageable in such scenarios as well. Likewise, it is easier to write or infer wrappers that extract individual results for pages of listings. Thus there are more opportunities to aggregate results from multiple sources and to let users slice and dice the retrieved results.

Typically, vertical search engines provide a deeper user experience in addition to search. For example, users expect to be able to perform transactions based on the search results (e.g., book an airline ticket) or see related information (e.g., crime rates in certain neighborhoods or product reviews). Therefore, vertical search engines are only viable when their domain has high commercial value. For that reason, virtual integration is *not* suitable in the context of a search engine. As we show in our earlier work [11], there are at least tens of millions of potentially useful forms and building and managing semantic mappings on such a scale can be very challenging. Further, forms cannot be classified into a small set of domains. Data on the web is about everything and boundaries of domains are not clearly definable. Hence, creating a mediated schema for the web would be an epic challenge (if at all possible), and would need to be done in over 100 languages.

In addition to the challenges of creating mediated schemas, a search engine employing the virtual approach would face significant challenges at query time. Since search engine users pose keyword queries, in order to use a virtual integration approach, at query-time the underlying system would have to identify the forms that are likely to have results relevant to the keyword query and then reformulate the keyword query into the most appropriate submissions over the chosen forms. To be effective, for each form, we would have to build models capable of identifying all possible search engine queries with results in the underlying content and be able to translate those queries into an appropriate query over the form. If the models are not precise, then we risk reformulating too many search engine queries that results in an unreasonable load on the underlying forms and a disappointing user experience.

### 3.2 The surfacing approach

The surfacing approach focuses on pre-computing the most relevant form submissions, i.e., queries, for all interesting HTML forms. The URLs resulting from these submissions are generated off-line and indexed in a search engine like any other HTML page. This approach is especially attractive to search engines, since it enables leveraging the existing infrastructure almost as-is and hence the seamless inclusion of deep-web pages.

In the surfacing approach, we do not have to solve the problem of building models that map keyword queries to corresponding forms. This problem is already solved by the underlying IR-index that is built by analyzing the contents of the pages resulting from form submissions. The challenge we face here is to pre-compute appropriate queries over the form, which involves two main technical problems. First, we have to identify values that are suitable for various form inputs. For select-menus, the values are already known, but for text inputs they need to be predicted. Second, we have to minimize the number of queries over each form so as to not pose an unreasonable load during off-line analysis. Specifically, a naive strategy like enumerating all possible queries can be fatal when dealing with forms with more than one input. In [12], we present algorithms that address these two problems. We have found that the number of URLs our algorithms generate is proportional to the size of the underlying database, rather than the number of possible queries.

It is interesting to point out that while the virtual-integration approach only answers queries that can be anticipated in advance, the surfacing approach can provide answers in a fortuitous manner. To illustrate, suppose you were to query a search engine for "SIGMOD Innovations Award MIT professor". Hypothetically, suppose that the only web source with that information is a deep-web site that lets you query for biographies of MIT professors by their departments. In a virtual integration system, it is likely that the form will be identified as one retrieving biographies given department names. Such a virtual integration system, even when it understands the form correctly, will not be able to retrieve the CSAIL department biographies page. However, the surfacing approach will have the CSAIL page indexed and would retrieve it correctly (based on matching the keywords to the page content), thereby identifying Mike Stonebraker to be the MIT professor with the SIGMOD Innovations Award.

The surfacing approach also leads to very reasonable loads on underlying form sites. User traffic is directed to deep-web content only when a user clicks on a URL generated from a form submission. At that point, the user has already deemed the URL to be relevant, which is a more accurate determination compared to a routing algorithm of the virtual approach. Note that when the user clicks on the URL, she will see fresh content. The search-engine web crawler contributes to some traffic during off-line analysis (before the pages are indexed) and index refresh (after the pages are indexed). However, the web-crawler traffic is reasonable and can be amortized over longer periods of time. Also, note that once the search-engine's index is seeded with good candidate deep-web URLs from a form-site, the web crawler will discover more content over time by pursuing links from deep-web pages. In [12], we show that we only impose light loads on the underlying sites, but are able to extract large portions of the underlying database.

While we have been able to show that surfacing can indeed make large quantities of deep-web content visible to search-engine users, it has some limitations. For example, surfacing cannot be applied to HTML forms that use the *POST* method. In a POST form, all form submissions have the same URL and the user query is embedded in the HTTP request rather than in the URL as it is done with GET requests. In most cases, the POST method is used when form submissions result in state changes to a back-end database, e.g., when buying a product or purchasing a ticket, which is anyway unsuitable for indexing in a search engine. However, there are cases where POST forms do include content that will be useful in a search-engine index.

Since the main goal of the surfacing approach is to provide users access to more content, an important question to consider is the impact of the surfaced content on the query stream, specifically, which types of queries is it really useful for? It is well known that the distribution of queries in search engines takes the form of a power law with a heavy tail, i.e., there are a large number of of rare queries. Our analysis shows that the impact of deep-web content is on the long tail of queries, thereby further validating the need for a domain-independent approach such as ours. We found that the pages surfaced by our system from the top 10,000 forms (ordered by the number of search engine queries they impacted) accounted for only 50% of deep-web results on Google.com, while even the top 100,000 forms only accounted for 85%. Thus, there are a very large number of forms that individually contribute to a small number of queries, but they together account for a large fraction of the queries. The reason the impact is on the long tail is that for the most popular queries, e.g., celebrity names or product searches, there are already a number of web sites with relevant content that are on the surface web. Further, the web sites in such popular domains have already been heavily optimized by search engine optimization (SEO) companies to make them easily accessible to search-engine crawlers. For such domains, the deep-web content provides very little added value. However, in domains with more rarely searched content, the surfacing of high-quality form-based sites can have a huge impact. Prominent examples include governmental and NGO portals. Such sites have structured content on rules and regulations, survey results, etc., but not the monetary resources to hire SEOs to make their content easily accessible.

## 4. THE ROLE OF SEMANTICS

In our work on surfacing deep-web content, we made very little use of the semantics of the forms we are crawling and their specific inputs. In contrast, semantics plays a crucial role in the virtual-integration approach to accessing the Deep Web. In this section we describe some of the areas where semantic analysis can play a significant role to extend the coverage of deep-web surfacing. In doing so, we raise several research challenges which can also be addressed outside of a search-engine company.

### 4.1 Semantics of form inputs

As mentioned earlier, we made very little use of semantics in deciding which values to enter into inputs of the forms we are surfacing. Conceivably, we could have designed a mediated schema with lists of values associated with different elements. If we found that a input in a form matches closely with an element in the mediated schema, the values for the element can be used to generate the queries we would pose to the form. To apply such an approach, we need to distinguish between two kinds of form inputs: search boxes and typed text boxes.

We found that the text boxes in a vast majority of forms have "search" boxes, i.e., they accept any keywords irrespective of their domain, and retrieve records that contain the search terms. To handle such search boxes, we generate candidate seed keywords by selecting the words that are most characteristic of the already indexed web pages from the form site. We then use an iterative probing approach to identify more keywords before finally selecting the ones that ensure diversity of result pages. It would have been difficult to map these search boxes to any specific set of elements in a mediated schema. Approaches similar to ours have be used by others in the context of extracting documents from text databases [1, 13].

The second type of text boxes are "typed" in that they do not accept arbitrary keywords. For such inputs, we believe it is more important to understand the data type of the specific input, because it can yield better coverage of the content behind the form and prevent us from posing meaningless queries to the form. Examples of such data types are US zip codes, city names, dates and prices. The important point here is that we do not need to know what the form is about (i.e., whether it retrieves store locations by zip-code or used-cars by make and zip-code) to surface content behind the form. All we need to know is that the text box accepts zip code values. We further showed that such common data types appear in large numbers of forms and can hence contribute significantly to surfacing. Our analysis indicates that as many as 6.7% of English forms in the US contain inputs of common types like zip codes, city names, prices, and dates. Fortunately, we also have preliminary results (reported in [12]) that suggest that one can identify such typed inputs with high accuracy.

Expanding on typed inputs, a natural question that arises is: what is a data type? For example, are names of car makes a data type? It turns out that form inputs that expect a value from a relatively small set of possibilities typically use select menus to guide the user. This creates a better user experience and simplifies the work of the back-end application. Other examples of such smaller domains are names of US state names, countries, cuisines and job types. Untyped text boxes are most often used when the space of possible values in very large (e.g., people names, ISBN values, product names).

## 4.2 Correlated inputs

Most prior approaches targeted at surfacing or extracting deep-web content have either considered forms with single input [1, 6, 13] or when considering multiple inputs queried them only one at a time [18] or ignored any dependencies between them [14]. However, most useful forms have multiple inputs and ignoring dependencies between different inputs can lead to ineffective surfacing results. In this section we outline two kinds of correlations between inputs that we consider especially important based on our experience – ranges and database selection.

**Ranges:** Forms often have pairs of inputs defining a range over a single numeric property – one input each to restrict the maximum and minimum values of the property in the retrieved results. For example, used-car sites allow for restricting listings by the price ranges, mileage ranges and year ranges. Our analysis indicates that as many as 20% of the English forms hosted in the US have input pairs that are likely to be ranges. Not recognizing such input pairs can lead to wasteful generation of URLs. Consider a form with two inputs, min-price and max-price, each with 10 values. Using the techniques described in [12], the two inputs would be tested independently, and might potentially lead to the erroneous identification of informative inputs. It it possible that as many as 120 URLs might be generated, many of which will be for invalid ranges. However, by identifying the correlation between the two inputs, we can generate the 10 URLs that each retrieve results in different price ranges.

To identify such correlations, we face two challenges. First, we must identify pairs of inputs that are very likely to correspond to ranges. Second, we must determine which pairs of values would be most effective to use for the ranges. Our initial efforts in this direction lead us to believe that large collections of forms can be mined to identify patterns (based on input names, their values, and position) for input pairs that constitute a range. Once ranges are identified, we have seen that even simple strategies for picking value pairs can significantly reduce the total numbers of URLs generated without a loss in coverage of the underlying content.

**Database selection:** A second common correlation pattern is database-selection. It consists typically of two inputs – one text box and one select-menu. The value in the text box poses keyword queries, while the select-menu identifies which underlying database the query is targeted at. For example, there are forms that let users search for movies, music, software, or games (indicated in the select menu) using keywords (in the single text box). The problem here is that the keywords that work well for software, e.g., "microsoft", are quite different from keywords for movies, music and games. In other words the best set of values for the text box varies with the value in the select menu. Generating different sets of keywords can be easily done, provided we can reliably identify such patterns in inputs, which is an open challenge.

We note that the canonical example of correlated inputs, namely, a pair of inputs that specify the make and model of cars (where the make restricts the possible models) is typically handled in a form by Javascript. Hence, by adding a Javascript emulator to the analysis of forms, one can identify such correlations easily.

## 5. ANALYSIS OF SURFACED CONTENT

Two problems areas that we believe are prime candidates for future deep-web research have to do with the interpretation of pages resulting from the surfacing.

## 5.1 Semantics and extraction

When structured data in the Deep Web is surfaced, the structure and hence the semantics of the data is lost. The loss in semantics is also a lost opportunity for query answering. For example, contrary to the fortuitous query-answering example we presented previously, suppose a user were to search for "used ford focus 1993". Suppose there is a surfaced used-car listing page for Honda Civics, which has a 1993 Honda Civic for sale, but with a remark "has better mileage than the Ford Focus". A simple IR index can very well consider such a surfaced web page a good result. Such a scenario can be avoided if the surfaced page had the annotation that the page was for used-car listings of Honda Civics and the search engine were able to exploit such annotations. Hence, the challenge here is to find the right kind of annotation that can be used by the IR-style index most effectively.

Going a step further, we pose a more ambitious challenge: is it possible to automatically extract relational data from surfaced deep-web pages? While there has been work done on generating wrappers and extracting tables from HTML pages, they have mostly been in the context of information extraction from web pages in general, rather than pages that were known to be generated by deep-web forms. Much of the work has focused on machine learning, where wrappers are inferred based on training data that is manually created by marking up desired structured components on different web pages on a given site. The huge variability in HTML layout make it a very challenging task in general. Hence, the challenge here is to extract rows of data from pages that were generated from deep-web sites where the inputs that were filled in order to generate the pages are known.

## 5.2 Coverage of the surfaced content

A question that is asked of any surfacing or database-extraction algorithm is: "what portion of the web site has been surfaced?" Ideally, we would like to quantify a candidate surfacing algorithm, with a statement of the form: with a probability of M% more than N% of the site's content has been exposed. However, this is an open research problem that does not yet have a satisfying solution. Most approaches to extracting deep-web content employ greedy algorithms that try to maximize coverage, but do not provide any guarantees.

It is important to note that a natural goal for a surfacing algorithm might be to minimize the number of surfaced pages while maximizing coverage. However, in [12], we argue that the surfacing goal for a search engine is different. Specifically, we argue that the web pages we surface must be good candidates for insertion into a search engine index. This implies that the pages we extract should neither have too many results on a single surfaced page nor too few. We present an algorithm that selects a surfacing scheme that tries to ensure such an indexability criterion while also minimizing the surfaced pages and maximizing coverage.

## 6. AGGREGATING STRUCTURED DATA ON THE WEB

The work on structured-data on the Web has focused mostly on providing users access to the data. However, our experience has shown that significant value can be obtained from analyzing collections of meta-data on the Web. Specifically, from the collections we have been working with (forms and HTML tables) we can extract several artifacts, such as: (1) a collection of forms (input names that appear together, values for select menus associated with input names), (2) a collection of several million schemata for tables, i.e., sets of column names that appear together, and (3) a collection of columns, each having values in the same domain (e.g., city names, zipcodes, car makes).

One of the challenges we pose is to use these artifacts to build a set of *semantic services* that are useful for many *other* tasks. Examples of such services include:

- Given an attribute name (and possibly values for its column or attribute names of surrounding columns) return a set of names that are often used as synonyms. In a sense, such a service is a component of a schema matching system whose goal is to help resolve heterogeneity between disparate schemata. A first version of such a service was described in [3].

- Given a name of an attribute, return a set of values for its column. An example of where such a service can be useful is to automatically fill out forms in order to surface deep-web content.

- Given an entity, return a set of possible properties (i.e, attributes and relationships) that may be associated with the entity. Such a service would be useful for information extraction tasks and for query expansion.

- Given a few attributes in a particular domain, return other attributes that database designers use for that domain (akin to a schema auto-complete). A first version of such a service was described in [3]. Such a service would be of general interest for database developers and in addition would help them choose attribute names that are more common and therefore avoid additional heterogeneity issues later.

We plan to make our data sets available to the community to foster research on these and other challenges.

## 7. REFERENCES

[1] L. Barbosa and J. Freire. Siphoning hidden-web data through keyword-based interfaces. In *SBBD*, 2004.
[2] M. J. Cafarella, A. Halevy, Y. Zhang, D. Z. Wang, and E. Wu. Uncovering the Relational Web. In *WebDB*, 2008.
[3] M. J. Cafarella, A. Halevy, Y. Zhang, D. Z. Wang, and E. Wu. WebTables: Exploring the Power of Tables on the Web. In *VLDB*, 2008.
[4] Cazoodle Apartment Search. http://apartments.cazoodle.com/.
[5] Every Classified. http://www.everyclassified.com/.
[6] L. Gravano, P. G. Ipeirotis, and M. Sahami. QProber: A system for automatic classification of hidden-web databases. *ACM Transactions on Information Systems*, 21(1):1–41, 2003.
[7] B. He and K. C.-C. Chang. Statistical Schema Matching across Web Query Interfaces. In *SIGMOD*, 2003.
[8] B. He, M. Patel, Z. Zhang, and K. C.-C. Chang. Accessing the Deep Web: A survey. *Communications of the ACM*, 50(5):95–101, 2007.
[9] P. G. Ipeirotis and L. Gravano. Distributed Search over the Hidden Web: Hierarchical Database Sampling and Selection. In *VLDB*, 2002.
[10] A. Y. Levy, A. Rajaraman, and J. J. Ordille. Querying Heterogeneous Information Sources Using Source Descriptions. In *VLDB*, 1996.
[11] J. Madhavan, S. Jeffery, S. Cohen, X. Dong, D. Ko, C. Yu, and A. Halevy. Web-scale Data Integration: You can only afford to Pay As You Go. In *CIDR*, 2007.
[12] J. Madhavan, D. Ko, L. Kot, V. Ganapathy, A. Rasmussen, and A. Halevy. Google's Deep-Web Crawl. *PVLDB*, 1(2):1241–1252, 2008.
[13] A. Ntoulas, P. Zerfos, and J. Cho. Downloading Textual Hidden Web Content through Keyword Queries. In *JCDL*, 2005.
[14] S. Raghavan and H. Garcia-Molina. Crawling the Hidden Web. In *VLDB*, 2001.
[15] A. D. Sarma, X. Dong, and A. Halevy. Bootstrapping pay-as-you-go data integration systems. In *SIGMOD*, 2008.
[16] M. Stonebraker. Byledge. Personal Communication.
[17] Trulia. http://www.trulia.com/.
[18] P. Wu, J.-R. Wen, H. Liu, and W.-Y. Ma. Query Selection Techniques for Efficient Crawling of Structured Web Sources. In *ICDE*, 2006.
[19] W. Wu, C. Yu, A. Doan, and W. Meng. An Interactive Clustering-based Approach to Integrating Source Query Interfaces on the Deep Web. In *SIGMOD*, 2004.